\documentclass[12pt,preprint]{aastex}
\begin{document}
\title{Evidence of Elevated X-Ray Absorption Before and During Major Flare Ejections in GRS 1915+105}
\author{Brian Punsly\altaffilmark{1}, J\'{e}r$\hat{\mathrm{o}}$me
Rodriguez\altaffilmark{2} and Sergei A. Trushkin \altaffilmark{3}}
\altaffiltext{1}{1415 Granvia Altamira, Palos Verdes Estates CA, USA
90274 and ICRANet, Piazza della Repubblica 10 Pescara 65100, Italy,
brian.punsly1@verizon.net or brian.punsly@comdev-usa.com}
\altaffiltext{2}{Laboratoire AIM, CEA/DSM-CNRS-Universit\'{e} Paris
Diderot, IRFU SAp, F-91191 Gif-sur-Yvette, France.}
\altaffiltext{3}{Special Astrophysical Observatory RAS, Nizhnij
Arkhyz, 369167, Russia}
\begin{abstract}
We present time resolved X-ray spectroscopy of the microquasar
GRS1915+105 with the MAXI observatory in order to study the
accretion state just before and during the ejections associated with
its major flares. Radio monitoring with the RATAN-600 radio
telescope from 4.8 - 11.2 GHz has revealed two large steep spectrum
major flares in the first eight months of 2013. Since, the RATAN
receives one measurement per day, we cannot determine the jet
forming time without more information. Fortunately, this is possible
since a distinct X-ray light curve signature that occurs preceding
and during major ejections has been determined in an earlier study.
The X-ray luminosity spikes to very high levels in the hours before
ejection then becomes variable (with a nearly equal X-ray luminosity
when averaged over the duration of the ejection) during a brief 3 to
8 hour ejection process. By comparing this X-ray behavior to MAXI
light curves, we can estimate the beginning and end of the ejection
episode of the strong 2013 flares to within $\sim$3 hours. Using
this estimate in conjunction with time resolved spectroscopy from
the data in the MAXI archives allows us to deduce that the X-ray
absorbing hydrogen column density increases significantly in the
hours preceding the ejections and remains elevated during the
ejections responsible for the major flares. This finding is
consistent with an out-flowing wind or enhanced accretion at high
latitudes.
\end{abstract}


\section{Introduction}
Astrophysical black holes reveal themselves to astronomers through
their interaction with adjacent gas. There are two observed aspects
to this phenomenon, the thermal glow from the viscously heated
accreting gas and the bipolar ejection of plasma that can propagate
very close to the speed of light. Both behaviors are seen in solar
mass black holes (microquasars) in the Milky Way as well as
supermassive black holes in active galactic nuclei (AGN). The
relationship between the accretion and jet formation has been
intensely studied (eg, \citet{fen04}, but is still not well
understood. Microquasars have a logistical advantage for astronomers
since significant evolution of the system can happen on short time
scales (ms to days). Thus, direct associations between accretion and
outflow diagnostics can be discerned, in principal, within a human
lifetime \citep{rod08,mil12}. By contrast, AGN vary on the time
scales of $\sim 1 - 10^{6}$ years. Thus, the chance of gathering
enough information to make statistically significant correlations
between accretion and outflow observable quantities in a human
lifetime is drastically diminished. Hence, the appeal of using
microquasars as time compressed laboratories to study black hole and
quasar physics. Substantial progress has been made in this regard
for GRS~1915+105 which is well known for ejecting radio ``blobs" out
to large distances with apparent superluminal speeds
\citep{mir94,fen99,dha00}. Punsly \& Rodriguez (2013a, PR13
hereafter) analyzed large RXTE data sets and demonstrated that
empirical relationships exist between the accretion states and major
flare ejections. In particular, the X-ray luminosity is highly
elevated in the last hours preceding major ejections and it is
correlated with the power required to eject the discrete blobs of
plasma at relativistic velocity. Secondly, the X-ray luminosity was
found to be highly variable during major flare ejections (MFEs,
hereafter), yet the time averaged X-ray luminosity during the
ejection event is correlated with that just before the MFE and is of
a similar (but perhaps a slightly lower) magnitude. However, the
ejections are brief and occur at unpredictable times. Thus, all the
relevant X-ray data are serendipitously gathered during wide field
monitoring. Therefore, we have no pointed X-ray observations near or
at the crucial instant of ejection and therefore no spectral data to
constrain the accretion state. Ironically, in this circumstance, the
rapid evolution of the microquasars impedes our understanding of the
black hole accretion system. Fortunately, the MAXI observatory
typically observes GRS~1915+105 every 1.5 hours and it has the
ability to gather spectral data from 1 keV - 20 keV \citep{mat09}.
For very bright states (as occurs during major ejections), single
epoch observations can provide a useful spectrum. For lower
luminosity states, with slower evolution, one can bin multiple
observations over a few hours to achieve sufficient signal to noise
for a useful spectrum.
\par Being thusly motivated to understand the X-ray spectral state
before and during major ejections, we have used the current
monitoring program with RATAN-600 radio telescope to find evidence
of major ejections (i.e., large radio fluxes and a steep radio
spectrum). Using a (limited) library of high temporal resolution
X-ray light curves of major ejections from PR13, we estimate the
time frame for flare ejection from the MAXI light curve. We
downloaded the MAXI spectral data adjacent to this epoch and look
for trends in the spectral fits and the raw number counts. Our
findings, based on the two largest radio flares in the first eight
months of 2013, are presented here.

\section{Radio Monitoring} We were granted radio monitoring time on
the RATAN-600 radio telescope at 4.8, 8.2 and 11.2 GHz for both the
first and second half on 2013. These observations have been carried
out in a current monitoring program of the microqusars with
RATAN-600 (eg \citet{tru08}). The cryogenic cooled receivers and
antenna were daily calibrated with secondary calibrators 3C286 and
PKS1345+12. High luminosity, steep spectrum, radio flares
originating from GRS 1915+105 have historically been associated with
superluminal ejections \citep{rus11}. Major flares begin as an
optically thick flux increase that becomes optically thin at lower
and lower frequency as they evolve over the next few hours. The
temporal spacing of our monitoring program -- one measurement per
day -- is far too large to estimate the time of the major ejection
that is associated with the radio flare. From PR13 and Figure 1,
ejection initiation and end times require temporal resolution 100
times finer, on the orders of 0.25 hours or less. However, strong
flares are luminous for at least 1 day, so it is adequate for
identifying major ejections \citep{rod99,mil05}.  We define major
flares as radio flares with a flux density $S_{\nu}>$ 100 mJy at
frequency 4.8 GHz, with a steep spectral index, $\alpha > 0.5$
($S_{\nu} \sim \nu^{-\alpha}$). We have identified 5 such major
flares in the 8 first months of 2013.

\begin{table}
\caption{RATAN - 600 Radio Data Near Major Flare Ejections}
{\footnotesize\begin{tabular}{ccccc} \tableline\rule{0mm}{3mm}
Date &  Flux Density & Flux Density & Flux Density & \\
MJD        &  4.8 GHz (mJy)     &  8.2 GHz  (mJy)        &  11.2 GHz (mJy) \\
\tableline \rule{0mm}{3mm}
        &       &  FLARE 1        &  \\
\tableline \rule{0mm}{3mm}
56312.349   &  $50 \pm 5$  & $50 \pm 7$ & $50 \pm 10$ \\
56313   &    & no observations &  \\
56314.349   &  $364 \pm 18 $  & $243 \pm 24 $ & $234 \pm 23 $ \\
\hline
  &       &  FLARE 2       &  \\
\hline
56349.253   &  $42 \pm 4$ & no data & $51 \pm 8$ \\
56350.250   &  $76 \pm 8$  & no data & $76 \pm 6$  \\
56351.248   &  $48 \pm 5$   & no data & $80 \pm 8 $ \\
56352.245   &  $207 \pm 11 $ & $166 \pm 17$ & $101 \pm 10$ \\
56353.242   &  $100 \pm 10 $ & $100 \pm 10$ & $100 \pm 10$ \\
\end{tabular}}
\end{table}
The correlation of X-ray luminosity with flare energy and power
shown in PR13 indicates that we should look at the strongest flares
in order to get adequate statistics in our MAXI data. Of these five
steep spectrum flares, two had $S_{\nu}>$ 200 mJy at 4.8 GHz. The
first (hereafter Flare 1) was detected on MJD 56314.349 and the
second (Flare 2) on MJD 56352.245. The observations adjacent to
these flares are summarized in Table 1. In the following, we
consider the MAXI spectral data for these flares. The other steep spectrum
radio flares that were not suitable for spectral study with MAXI are
discussed in the Appendix.

\section{MAXI Observations of Major Flares}
We downloaded the data from the MAXI pipeline around the times of
the strong major radio flares listed in Table 1. We aim to
accomplish two goals with these data. First is to estimate the dates
that the major ejections occurred. Secondly, we implement time
resolved MAXI spectroscopy to determine the evolution of the X-ray
spectrum preceding and during major ejections.

\subsection{Estimating the Date of the Ejection Episode of Flare 2}
According to Table 1, the detection on MJD 56352.245
indicates a strong flare that has evolved far past the launch point,
since it is very steep spectrum. This flare is the primary topic of
this study since it has the most complete MAXI spectral coverage.
\begin{figure}
\begin{center}
\includegraphics[width=73 mm, angle= 0]{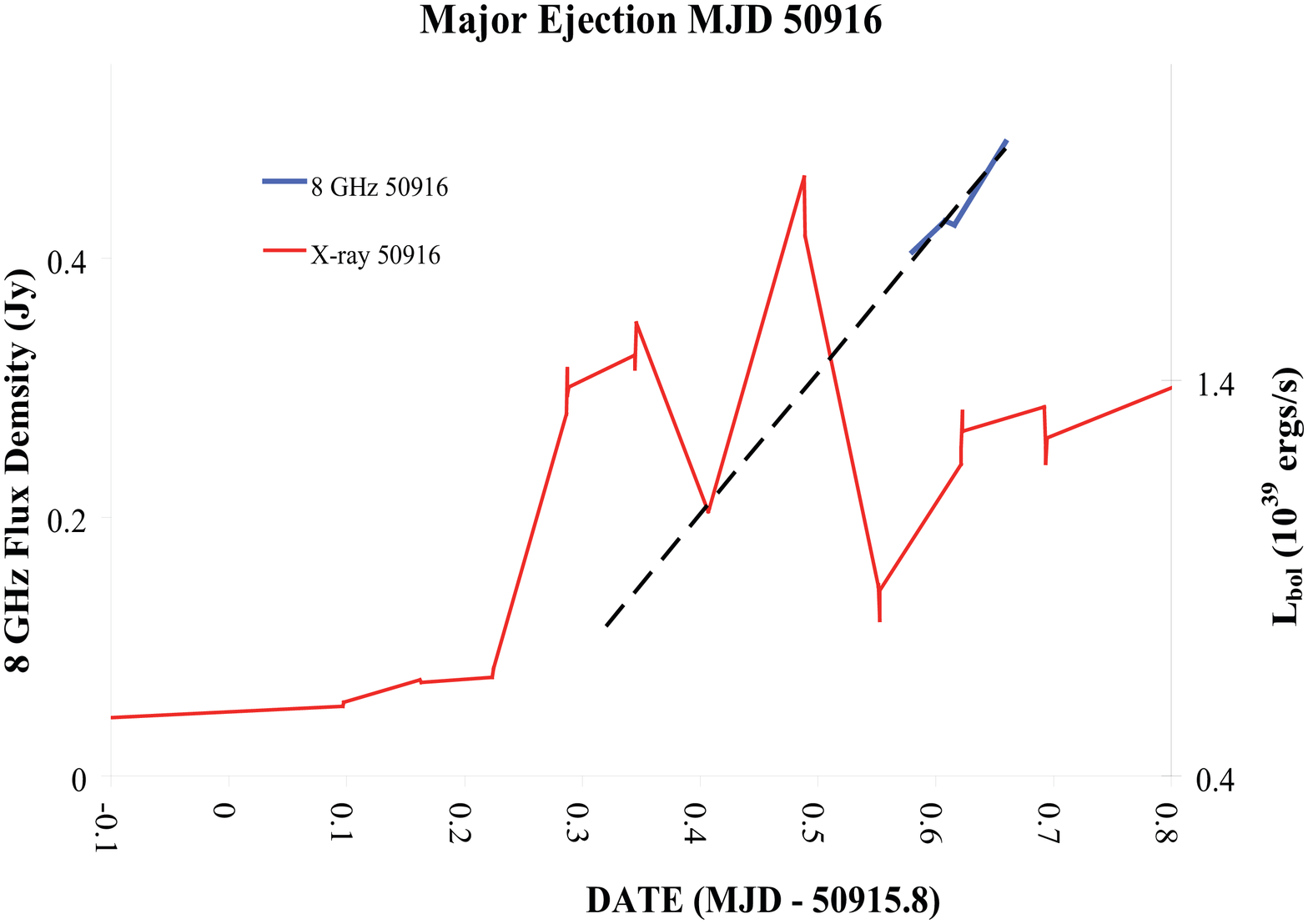}
\includegraphics[width=73 mm, angle= 0]{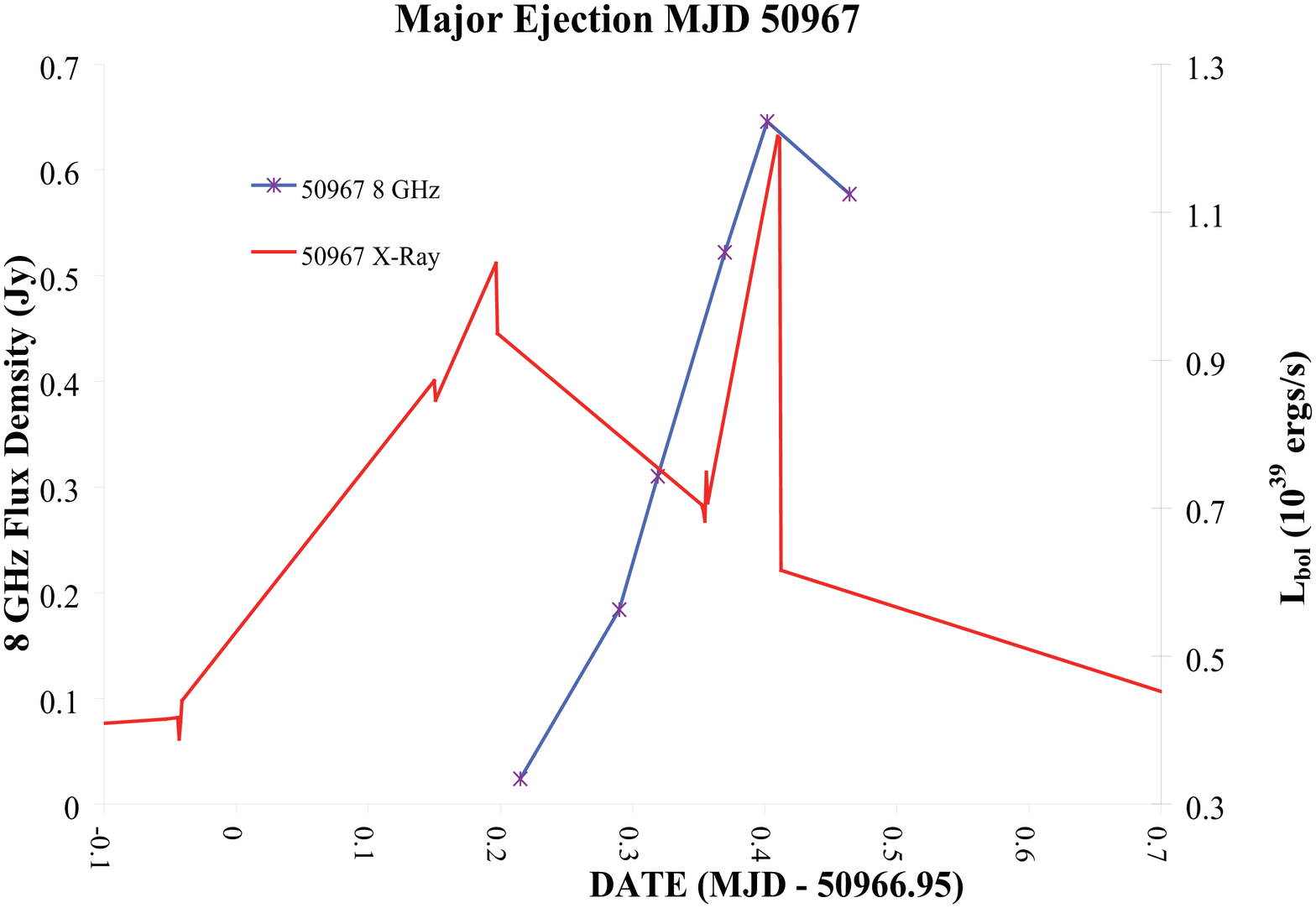}
\includegraphics[width=100 mm, angle= 0]{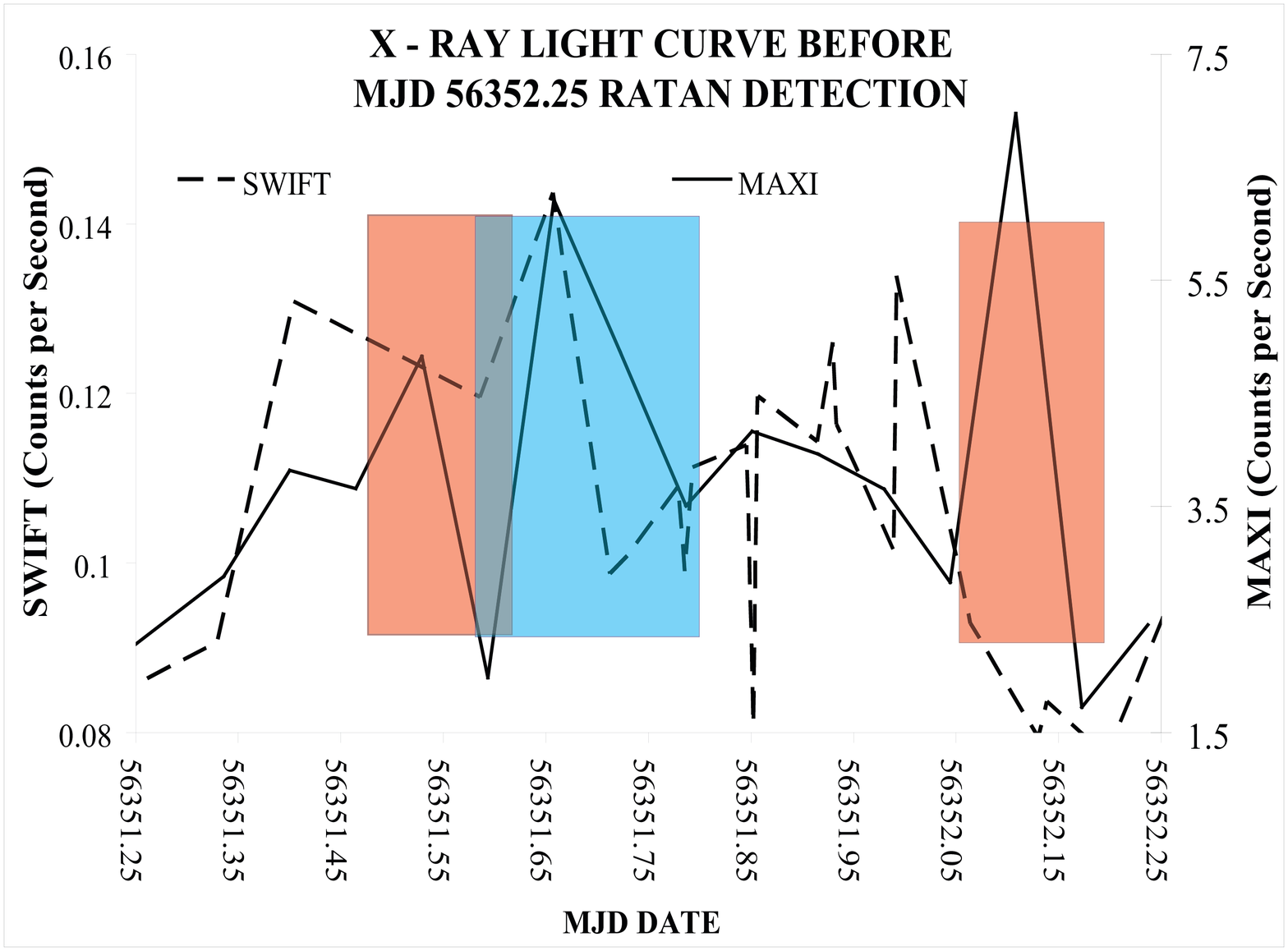}
\includegraphics[width=73 mm, angle= 0]{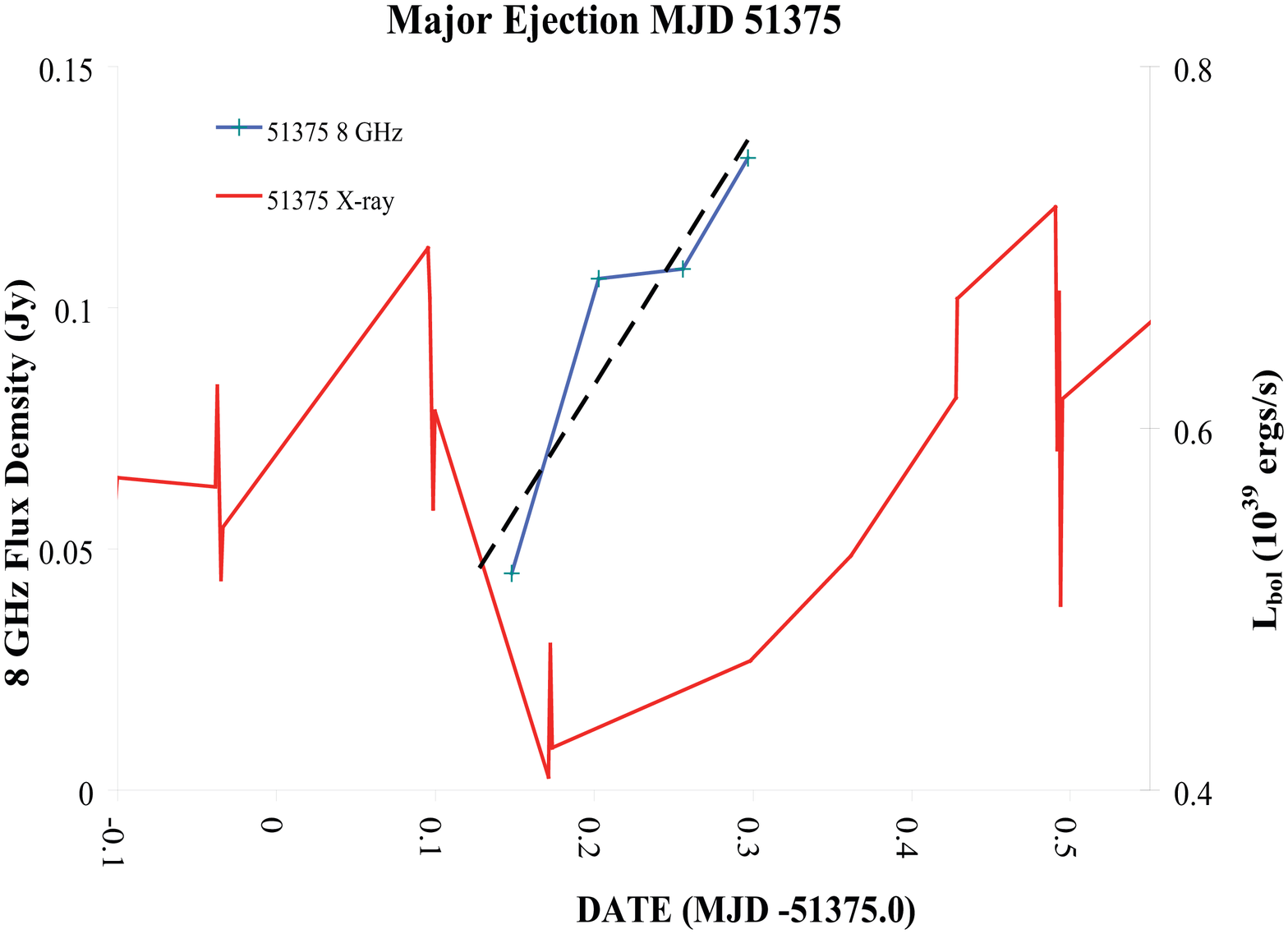}
\includegraphics[width=73 mm, angle= 0]{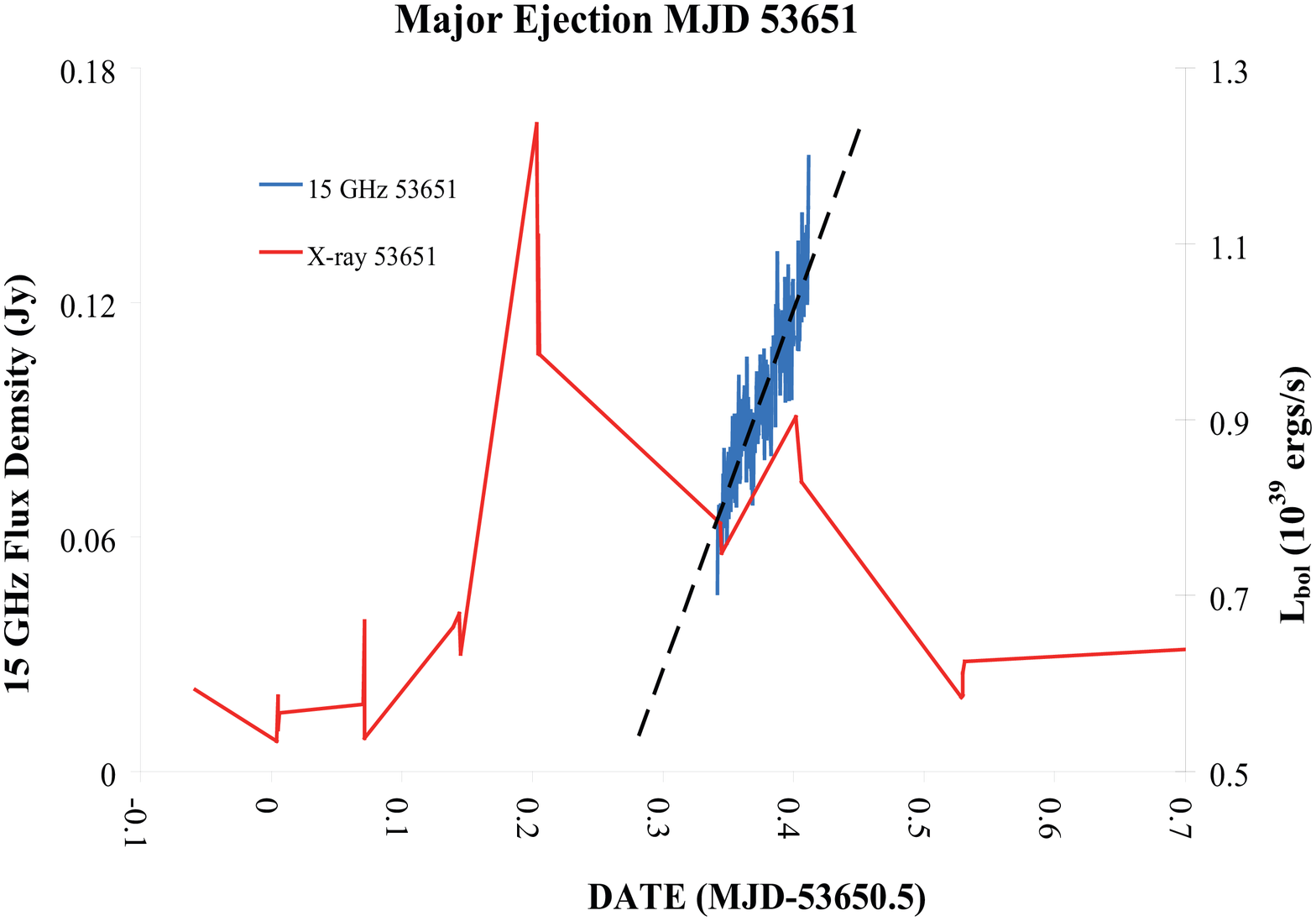}
\end{center}
\caption{A montage comprised of previously published X-ray
(RXTE/ASM) light curves associated with major flares in the four
corners of the figure with the MAXI 2 keV - 20 keV and SWIFT/BAT 10
keV - 50 keV light curves before the 56352.245 radio flare plotted
in the middle frame. The radio data (blue) and the associated trend
line (dashed black) for the four historical light curves are plotted
over the estimated interval of the MFE from PR13 and
\citet{pun15}. In the central frame, the shaded red (blue) area
represents the minimal (maximal) ejection duration from the sample
in PR13.}
\end{figure}
Without a time-line set by the beginning and end of the MFE episode,
time resolved spectroscopy cannot be connected to a causal physical
chain of events associated with the ejection process. Thus,
establishing the date of the ejection episode is crucial to the
following analysis.
\subsubsection{X-ray Diagnostics of Ejection} For
sufficiently strong flares, the results of PR13 and \citet{pun15}
can be used to define a clear X-ray light curve signature before and
during major ejections. To this end, we utilize transcribed light
curves from these references in Figure 1. In particular, the 4 light
curves (based on the bolometric X-ray luminosity,
$L_{\mathrm{bol}}$) for which we not only have X-ray luminosity
estimates in the final hours preceding the ejection, but also at
least two time separated luminosity estimates during the ejection
proper. Inspection of the four light curves indicate the following
behavior:
\begin{enumerate}
\item A large X-ray luminosity spike just before the MFE is launched.
\item During the ejection proper, the X-ray luminosity is highly variable.
\item There is always at least one dip in the X-ray luminosity
during the MFE proper, well below the pre-launch
value.
\item From Figure 1, PR 13 and \citet{pun15}, the time averaged X-ray
luminosity during launch is highly correlated with the luminosity
preceding the launch with a similar (but perhaps slightly smaller)
luminosity.
\item There can be large spikes in the light curve during the MFE which can exceed the
X-ray luminosity before the launch. These spikes can occur during
the ejection proper (see the MJD 50916 light curve in Figure 1) or
immediately after the ejection episode has terminated (see the MJD
50967 light curve in Figure 1).
\item From the data in PR 13, the widest possible range of ejection episode
durations for flares with 200 mJy to 400 mJy flux density at 4.8 GHz
is 0.13 days to 0.20 days.
\end{enumerate}
The other light curves associated with major flares in PR13 are
consistent with the behaviors of the four light curves in Figure 1
that are listed above. However, there is insufficient data sampling
during the launch proper to demonstrate more than points 1 and 6
above. Hence, we need to rely heavily on this small subsample. In
general, the 6 items above should be sufficient to greatly restrict
the plausible ejection epochs for individual flares from MAXI light
curves that are typically sampled every 0.06 days.
\subsubsection{Prominent Features in the X-ray Light Curve} We implement the 6 X-ray diagnostics of flare
ejection to estimate the epoch of ejection corresponding to the
Flare 2 in Table 1. This is actually a fairly complicated light
curve. There are three spikes in the MAXI count rate between the
flare non-detection by RATAN-600 on MJD 56351.248 (dropping the MJD,
hereafter) and the detection on 56352.245: 56351.54, 56351.66 and
56352.13. The red (blue) shaded areas in Figure 1 correspond to the
minimum (maximum) plausible ejection duration from point 6, 0.13
(0.20) days. We consider the prominent features of X-ray light curve in the context of the 6 points
above:
\begin{itemize}
\item \textbf{Dip (56351.59)}: The abrupt dip following a large spike in the X-ray light curve at 56351.59 is typical of X-ray light curves during
ejection, never before ejection (see point 3 above). Thus, we
conclude that the ejection begins before the dip and ends after the
dip.

\item \textbf{Spike 1 (56351.54)}: Recall the pattern described by points 1 - 3 above: a large
spike in the X-ray light curve precedes ejection and appears just
before large fluctuations in the light curve that typically show a
very large dip during the ejection episode. Spike 1 precedes the
profound dip at 56351.59 and fits this pattern. We conclude that
Spike 1 is either just before or during the ejection. The data and
our method are insufficient to discriminate between the
possibilities. Due to gaps in the MAXI coverage, one cannot assume
that the actual local maximum of the light curve has been captured
by the observations. Formally, one can only restrict the actual
light curve local maximum (Local Maximum 1):
$56351.47<\mathrm{Local\; Maximum\;1} < 56351.58$.

\item \textbf{Spike 2 (56351.66)}: Comparison to other light curves in Figure
1 indicates that Spike 2 is either during the ejection episode or
immediately following ejection (point 5 above). Again, the data and
our method are insufficient to discriminate between the
possibilities. Due to temporal gaps in the MAXI coverage, as above, one can only restrict the
associated local maximum in the light curve (Local Maximum 2) by
$56351.60<\mathrm{Local\; Maximum\;2} < 56351.72$.
\item \textbf{Spike 3 (56352.13)}: Spike 3 is only 0.11 days from the well
developed flare observed by RATAN, this is too close in time to be
probable from an evolutionary standpoint (PR13). Furthermore, the
red shading shows that it is too brief to be responsible for jet
launch since the MAXI counts are already low before 0.13 days have
elapsed (point 6) and remain low up until the RATAN observation. By
contrast, according to point 4, we expect a high average X-ray
luminosity during the ejection. Finally, the suppressed SWIFT counts
(from 10 keV - 50 keV) shows that the X-ray spectrum is very steep.
This phenomenon of very luminous variable steep spectrum X-ray
states \emph{after} major ejections is well documented
\citep{dha00,nam06,tru07}. Thus, based on these three reasons, we
conclude that Spike 3 is not associated with flare launch.
\end{itemize}
\subsubsection{Ejection Episode Start Time} From the first and second bullets in subsection 3.1.2, we conclude that
the ejection begins before the dip at 56351.59 and ends after the
dip. Equivalently, the latest possible start date is 56351.58. From
point 1 in subsection 3.1.1, we expect a large spike in the X-ray
luminosity before the MFE. From bullet 2 in subsection 3.1.2 this
occurs at $56351.47<\mathrm{Local\; Maximum\;1} < 56351.58$. Thus,
we conclude that the major ejection begins between 56351.47 and
56351.58.

\subsubsection{Ejection Episode End Time}
Applying the historical range of ejection durations from point 6 in subsetion 3.1.1, of 0.13 days to 0.20 days to the range of
start times in subsection 3.1.3 will bound the end of the MFE. In
summary, we estimate that the major ejection begins between 56351.47
and 56351.58 and ends between 56351.60 and 56351.78.

\subsection{Time Resolved Spectroscopy of Flare 2} Based on our estimates of the ejection episodes above, we use the
MAXI spectral data to look for spectral evolution before and during
ejection (i.e., we are looking for evidence of the physical
mechanism responsible for ejection). Since the sensitivity of MAXI
is low (low effective area), binning of the data over several times
the temporal resolution is often required in order to constrain our
spectral fits. This greatly reduces the time resolution except at
strong peaks. The results obtained from the spectral fits with a
model consisting of an absorbed powerlaw ({\tt{tbabs*powerlaw}} in
{\tt{XSPEC}} terminology) with abundances obtained from
\citet{wil00} to the MAXI data from 1 keV to 20 keV are listed in
Table 2. The first column represents the range of dates in each bin.
The second column is the fitted column density of hydrogen, $N_{H}$,
responsible for the effects of absorption in the spectrum. The third
column is the photon index of the power law fit to the data,
$\Gamma$. The next two columns are the absorbed flux and the
unabsorbed (intrinsic) flux that de-convolves the effects of
$N_{H}$. The last column is the reduced $\chi^{2}$ figure of merit
of the fit. The errors on the fitted parameters are at 90\%
confidence and the errors on the flux are at 68\% confidence (1
$\sigma$).
\begin{table}
\caption{Parametric Fits to MAXI Data Near Major Flare Ejections}
{\footnotesize\begin{tabular}{ccccccc} \tableline\rule{0mm}{3mm}
Date &  $N_{H}$ & $\Gamma$ & Observed Flux  & Intrinsic Flux & Reduced\\
 &   &  & 1 - 10 keV  & 1 - 10 keV & $\chi^{2}$(dof)\\
MJD &  $10^{22} \mathrm{cm}^{-2}$& &  $10^{-8} \mathrm{ergs}/\mathrm{sec-}\mathrm{cm}^{2}$ & $10^{-8} \mathrm{ergs}/\mathrm{sec -} \mathrm{cm}^{2}$ & \\
\tableline \rule{0mm}{3mm}
$56312.32 \pm 0.16$ & $8.2^{+2.0}_{-1.7}$ &$2.7 \pm 0.2$ & $1.56 \pm 0.06 $ & $5.05 \pm 0.19$ & 1.40(79)\\
$56312.64 \pm 0.10$   &  $10.5^{+3.6}_{-3.0}$  & $3.1^{+0.4}_{-0.3}$  & $ 1.68 \pm 0.15$ & $9.50 \pm 0.85$ & 1.20(42)\\
$56312.84\pm 0.04$   &  $10.9^{+2.5}_{-2.1}$  & $3.0 \pm 0.3$ & $3.10 \pm 0.20$ & $15.10 \pm 0.80$ & 0.90(66)\\
$56313.06\pm 0.01$   &  $16.7^{+4.4}_{-3.5}$  & $3.2 ^{+0.4}_{-0.3}$ & $5.20 \pm 0.40$ & $42.80 \pm 5.00$& 1.35(41)\\
$56313.16\pm 0.04$   &  $9.7^{+4.1}_{-3.1}$  & $2.7^{+0.4}_{-0.3}$ & $3.20 \pm 0.30$ & $12.40 \pm 1.60 $ & 0.90(27)\\
\hline \hline
$56350.85\pm 0.20$   &  $6.4^{+1.7}_{-1.4}$  & $2.4 \pm 0.2$ & $ 1.31 \pm 0.07 $ & $3.36 \pm 0.18 $ & 0.88(75)\\
$56351.20 \pm 0.14$   &  $9.3^{+2.0}_{-1.7}$  & $2.8 \pm 0.2$  & $ 1.63 \pm 0.09 $& $ 6.28 \pm 0.37 $ & 1.27(68)\\
$56351.46\pm 0.07$   &  $10.1^{+2.3}_{-2.0}$  & $2.8 \pm 0.2$ & $ 3.00 \pm 0.20 $ & $ 12.20 \pm 0.82 $ & 0.90(62)\\
$56351.54\pm 0.01$   &  $11.2^{+3.2}_{-2.7}$  & $2.9 \pm 0.3 $ & $ 3.90 \pm 0.30 $ & $ 18.00 \pm 1.39 $& 0.80(43)\\
$56351.59\pm 0.01$   &  $9.7^{+4.7}_{-3.4}$  & $3.2^{+0.6}_{-0.5}$ & $ 1.4 \pm 0.30 $ & $ 8.70 \pm 2.60 $ & 0.90(19)\\
$56351.66\pm 0.01$   &  $14.2^{+3.0}_{-2.5}$  & $3.2 \pm 0.3$ & $ 4.70 \pm 0.40 $ & $ 33.40 \pm 2.84 $ & 1.00(55)\\
$56351.83\pm 0.13$   &  $11.0^{+1.6}_{-1.4}$  & $2.8^{+0.2}_{-0.1}$ & $ 2.2 \pm 0.10 $ & $ 14.30 \pm 0.65 $& 0.99(117)\\
\end{tabular}}
\end{table}
There are some evident trends in Table 2 for the 56352.245 flare.
Obviously, the luminosity increases before the ejection. But, we
also see $\Gamma$ steepen and this steep spectrum seems to continue
during the luminosity variations that occur during the ejection. The
uncertainty in the photon index in Table 2 makes this phenomenon
less than statistically significant. However, we also see an
increase $N_{H}$ with more statistical significance and we explore
this below in Figure 2.
\begin{figure}
\begin{center}
\includegraphics[width=130 mm, angle= 0]{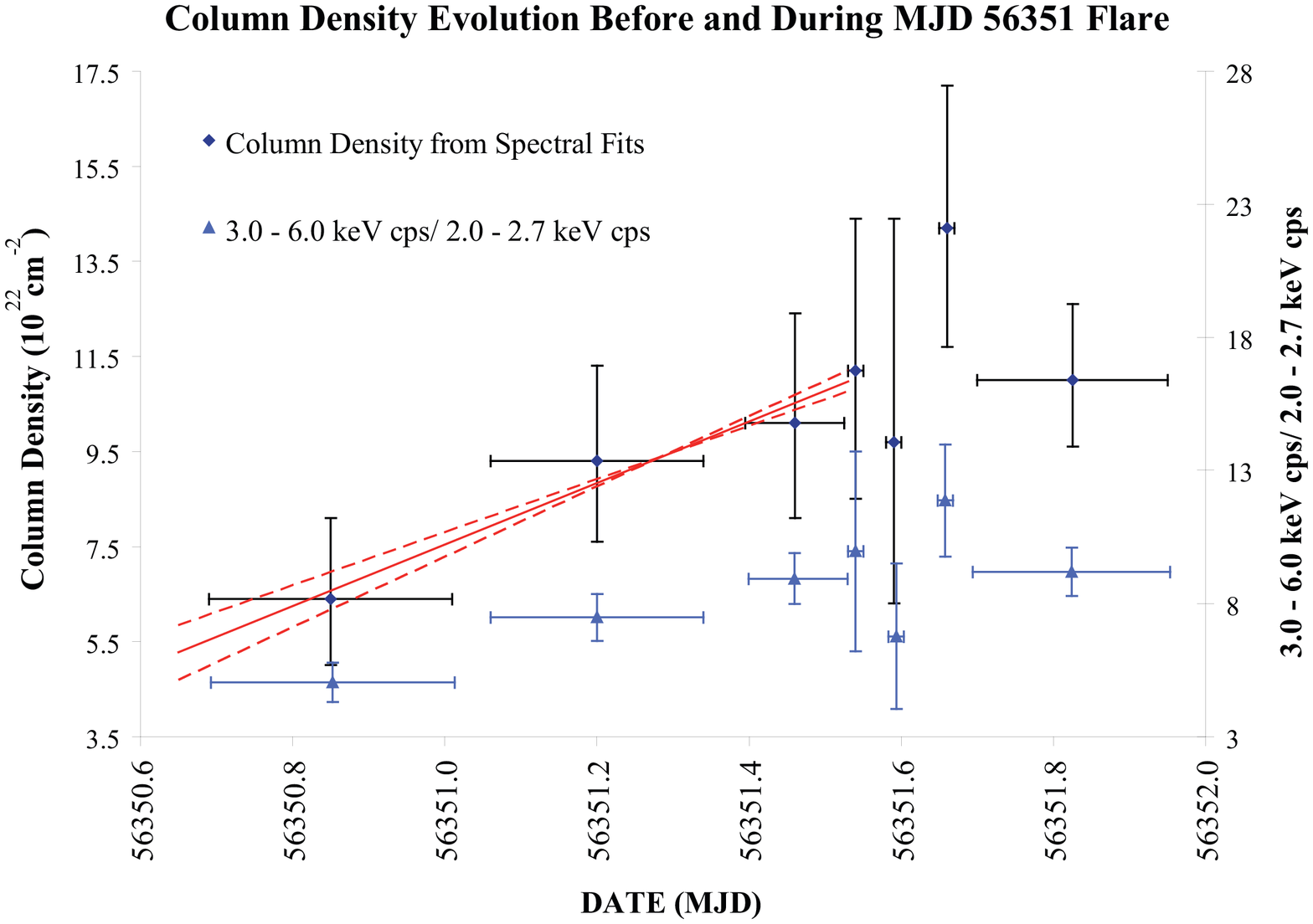}
\end{center}
\caption{The fitted $N_{H}$ from 56350.85 to 56351.83 from Table 2
is plotted in black. The fit to the data by the method of least
squares with uncertainty in both variables form 56350.85 to 56351.54
is plotted as the solid red line. The maximum and minimum slopes
associated with 1$\sigma$ uncertainty are the dashed red lines. The
blue triangles are a nonparametric measure of absorption (the
hardness in the soft band), the MAXI 3.0 keV - 6.0 keV cps divided
by the 2.0 keV - 2.7 keV cps.}
\end{figure}
The data in Figure 2 (from Table 2) is suggestive of a steady
increase in $N_{H}$ during the 15 hours preceding the ejection.
However, the errors are large and it does not warrant such a strong
statement without a statistical analysis. To this end note that the
solid red line in Figure 2 is the fit to the data from 56350.85 up
to 56351.54 using the method of least squares with uncertainty in
both variables \citep{ree89}. Equations (1) and (2) are the fits
from 56350.85 up to dates 56351.54 and 56351.66, respectively:
\begin{eqnarray}
&& N_{H} = (6.48 \pm 0.92)(\mathrm{DATE} - 56350.65) + (5.27 \pm
0.57) \;,\;\mathrm{DATE} \leq 56351.54 \;,\\
&& N_{H} = (7.49 \pm 1.67)(\mathrm{DATE} - 56350.65) + (4.83 \pm
1.20) \;,\;\mathrm{DATE} \leq 56351.66 \;.
\end{eqnarray}
The dashed lines in Figure 2 indicate the maximum and minimum slopes
consistent with the data at the 1$\sigma$ level based on Equation
(1). The slope and the corresponding uncertainty in Equation (1)
indicates an increase in $N_{H}$ at the $\sim 6.5\sigma$ level
preceding the flare.
\par Even thought this result is statistically robust, we investigate
if it is a manifestation of the degeneracy between $\Gamma$ and
$N_{H}$ in the parametric spectral fits. Within a power law model, a
modest low energy flux is only consistent with a large $\Gamma$ if
$N_{H}$ is large as well. This is a consequence of the power law
assumption. To this end, we use a nonparametric diagnostic, the
ratio of counts per second (cps) in a MAXI bin from 3.0 keV to 6.0
keV (the spectral peak) to cps in the lowest MAXI energy bin 2.0 keV
to 2.7 keV. Of course, the lowest energy bin is the most susceptible
to the effects of absorption and the cps will be low and our
statistics will suffer accordingly. Thus, considerable binning in
the time domain is required to improve the statistics. The advantage
of this nonparametric method of analysis is that it does not depend
on $\Gamma$ and the power law assumption. This removes the
parametric degeneracy and replaces it with a ratio that is purely an
empirical number. The blue data points indicate that the 2.0 keV to
2.7 keV cps are ever more suppressed relative to the 3.0 keV to 6.0
keV cps as the ejection is approached. This nonparametric analysis
supports the power law based deduction that $N_{H}$ increases before
the major ejection.
\subsection{MAXI observations of the Major Flare 1}
Even though Flare 1 is larger than Flare 2, the MAXI data suffers
from some inopportune gaps in temporal coverage and is inferior for
our purposes.
\begin{figure}
\begin{center}
\includegraphics[width=130 mm, angle= 0]{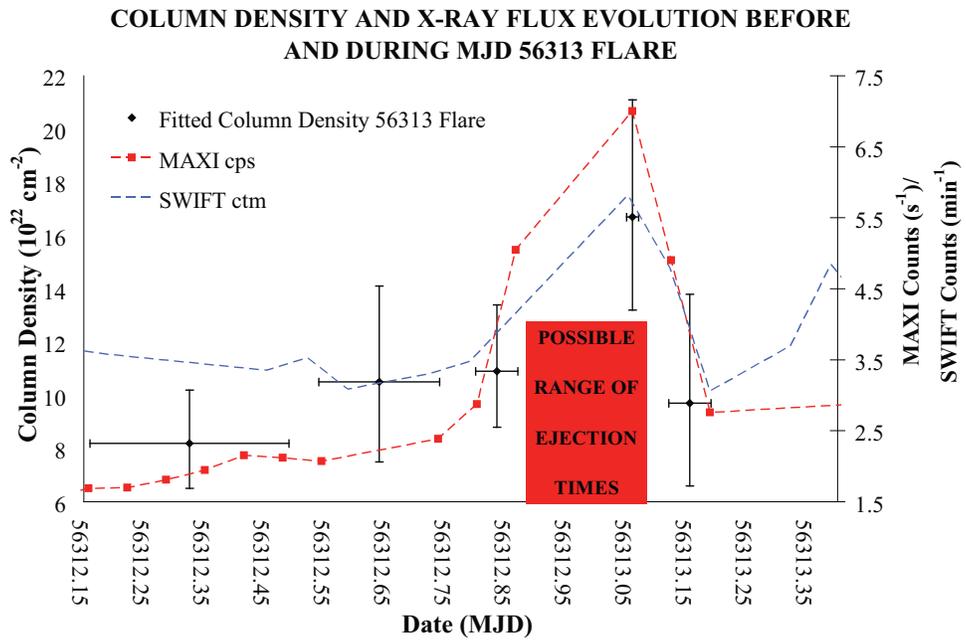}
\end{center}
\caption{The MAXI light curve before Flare 1 is plotted in cps and
the SWIFT light curve is plotted in ctm (counts per minute). The
fitted column densities before the flare are taken from Table 2.}
\end{figure}
Figure 3 plots $N_{H}$ from the spectral fits in Table 2, the MAXI 2
keV - 20 keV light curve and the SWIFT 10 keV - 50 keV light curve.
The estimated range of possible ejection times, 56313.86 - 56314.06
are based on the 6 criteria listed above in Section 3. There was a
large gap in MAXI coverage that envelopes most of the ejection
episode. The statistics are poor because most of the luminous bins
are lost due to the MAXI gap near ejection (when the fluxes were the
highest). Furthermore, the X-ray cps decrease very rapidly after
achieving a peak near 56314.0.  Thus, large bins are required to get
adequate statistics. The combination of wide ranges of dates in the
bins and what appears to be a steep increase in $N_{H}$ after
56313.8 results in an insufficient number of bins to resolve the
increase. Thus, we cannot claim a large statistically significant
increase in $N_{H}$ preceding launch. The increase in $N_{H}$ in
Figure 3 before the MFE episode is consistent with the increase seen
prior to the MFE that is responsible for Flare 2.
\section{Conclusion} In this article, we examined time resolved
MAXI X-ray spectroscopy of two major flares of GRS 1915+105 found
with RATAN-600 radio monitoring during the first 8 months on 2013.
The time resolved spectroscopy indicates that $N_{H}$ increases as
the X-ray luminosity increases before a major ejection and appears
to remain elevated until the end of the ejection episode. This
behavior was seen convincingly for Flare 2 and appears to be
consistent with poorer data from Flare 1.
\par For the flares considered here, we determined that $N_{H}$ increased by $\approx
5-8 \times 10^{22} \mathrm{cm}^{-2}$ over a time frame that begins a
few hours before the initiation of the ejection and appears to
extend well into the ejection episode. The increase of $N_{H}$ can
be cast in the light of the increase in the intrinsic absorption due
to the column density located within the GRS1915+105 binary system,
$N_{H}(\mathrm{intrinsic})$. There is a significant component of the
absorbing column due to the line of sight across the Galaxy, an
extrinsic component, $N_{H}(\mathrm{extrinsic})$. It was estimated
in \citet{cha04} that $N_{H}(\mathrm{extrinsic}) \approx 3.5 \times
10^{22} \mathrm{cm}^{-2}$. The discussion in \citet{bel97} indicates
a slightly higher stable interstellar absorption column,
$N_{H}(\mathrm{extrinsic}) \approx 4.5 \times 10^{22}
\mathrm{cm}^{-2}$. From Table 2, this means that
$N_{H}(\mathrm{intrinsic}) \approx 0.0 - 1.5 \times 10^{22}
\mathrm{cm}^{-2}$ 12 - 15 hours before the ejection episode and
increases by a factor $>10$ by the time of the ejection.

\par The increase in $N_{H}$ has two obvious plausible physical interpretations
\begin{itemize}
\item The increase represents accreting gas at relatively high latitudes, since
the line of sight is estimated at $65^{\circ} - 70^{\circ}$ to the
accretion disk normal from superluminal ejection kinematics
\citep{mir94,fen99}. The additional high latitude accretion
component is then a part of the triggering (and sustaining)
mechanism of the major ejections.
\item The enhanced absorbing column represents an out-flowing
wind from the accretion disk that begins before a major ejection and
continues during the ejection.
\end{itemize}
There is not enough data to conjecture deeply on these alternatives
from a physical perspective. However, the first alternative is
consistent with 3-D numerical simulations of accretion onto rotating
black holes. A small class of simulations were found to be
consistent with the correlation of the power of the ejection and the
elevated  X-ray luminosity. Namely, simulations that produced an
ergospheric disk jet about a black hole with spin parameter
$a>0.984$. In the simulations, the jet power and the X-ray
luminosity can be strong only during episodes of elevated accretion
\citep{pun14}. In order to understand if the behavior of $N_{H}$
seen here is endemic to large flares as opposed to anecdotal, we
plan on continuing the RATAN monitoring to find more major flares.
In particular, because of the correlation of radio luminosity before
and during ejections and the X-ray flux found in PR13, stronger
radio flares will have larger corresponding X-ray fluxes around the
ejection episode. Thus, an extremely powerful flare (i.e., larger
than 500 mJy at 4.8 GHz) will have large MAXI cps and smaller bins
can be used thereby improving the time resolution and the
statistics.
\section{Appendix: Other Steep Spectrum Radio Flares in 2013}
In this Appendix, we describe the other steep spectrum radio flares
from the first 8 months of 2013 with $>$ 100 mJy flux density at 4.8
GHz, $S_{\nu}(4.8 \mathrm{\; GHz}) > 100$ mJy. None of these flares
have suitable X-ray data for time resolved spectroscopy.
\begin{itemize}
\item 56439.87 is the date of the RATAN-600 detection of the first of a
series of flares that culminated in the large flare on 56352.25 (Flare
2) that was described at length in the text. RATAN-600 measured
$S_{\nu}(4.8 \mathrm{\; GHz}) \approx 140$ mJy and the spectrum was
steep. Unfortunately, the MAXI scan stopped on GRS 1915+105, and did
not complete when MAXI turned off its high voltage in order to
protect X-ray detectors in a high radiation region. The process
removed the data since the scan did not complete (Tatehiro private
communication 2013).
\item 56445.25 is the date of the RATAN-600 detection of
$S_{\nu}(4.8 \mathrm{\; GHz}) \approx 120$ mJy and the spectrum was
steep. Unfortunately, MAXI did not cover GRS 1915 during the flare
(Tatehiro private communication 2013).
\item On 564543.97, RATAN-600 measured $S_{\nu}(4.8 \mathrm{\; GHz})= 175 \pm
15$ mJy and the spectrum was steep. The 2 keV - 20 keV MAXI light curve that was sampled every 0.06 days was in the range of 1.7 cps to 2.8
cps near the time of flare launch (the local maximum). This is less
than half of what we found near the ejection period for Flare 2 in
Figure 1 and about 1/3 of the count rate for Flare 1 in Figure 3
with the same bin size in the time domain. Thus, the count rate is
too low for time resolved spectroscopy as the uncertainty in the
fits would be enormous due to the low number statistics. This could
be compensated for by approximately doubling the bin size in the
time domain, but this would not provide meaningful time resolution.
This modest flare highlights the importance of isolating the
strongest radio flares for MAXI time resolved spectroscopy.
\item The flare on 56487.87 detected by RATAN-600 was not formally
steep: $S_{\nu}(4.8 \mathrm{\; GHz})= 163 \pm 16$ mJy, $S_{\nu}(8.2
\mathrm{\; GHz})= 160 \pm 16$ mJy and $S_{\nu}(11.2 \mathrm{\;
GHz})= 107 \pm 11$ mJy. It is steep spectrum from 8.2 GHz to 11.2 GHz, but
flat spectrum between 4.8 GHz and 8.2 GHz. It seems likely that
$S_{\nu}(4.8 \mathrm{\; GHz})$ would be larger than 100 mJy and
steep spectrum a few hours later if the radio emitting plasma were in
a state of rapid expansion. Unfortunately, like the 56439.87 radio
flare, the MAXI scan stopped on GRS 1915+105, and did not complete.
The process removed the data since the scan did not complete
(Tatehiro private communication 2013).
\end{itemize}

\begin{acknowledgements}
This research has made use of MAXI data provided by RIKEN, JAXA and
the MAXI team. BP would like to thank Tatehiro Mihara of the MAXI
team for helping with the data acquisition and analysis. JR
acknowledge funding support from the French Research National
Agency: CHAOS project ANR-12-BS05-0009
(http://www.chaos-project.fr). JR also acknowledges  financial
support from the UnivEarthS Labex program of Sorbonne Paris Cité
(ANR-10-LABX-0023 and ANR-11-IDEX-0005-02). SAT is thankful to
Russian Foundation for Base Research for support (Grant
N12-02-00812). We were also very fortunate to have a referee who
offered many useful comments.
\end{acknowledgements}

\end{document}